\documentclass[twocolumn]{revtex4}

\tighten

\include{epsf}

\begin{document}

\title{\textsc{The Modern FPGA as Discriminator, TDC and ADC} }

\author{ Gary~S.~Varner}
\affiliation{Department of Physics \& Astronomy\\
University of Hawaii at Manoa \\
2505 Correa Road  Honolulu, HI 96822  USA}

\pagestyle{headings}

\begin{abstract}
Recent generations of Field Programmable Gate Arrays (FPGAs) have
become indispensable tools for complex state machine control and
signal processing, and now routinely incorporate CPU cores to allow
execution of user software code.  At the same time, their exceptional
performance permits low-power implementation of functionality
previously the exclusive domain of dedicated analog electronics.
Specific examples presented here use FPGAs as discriminator,
time-to-digital (TDC) and analog-to-digital converter (ADC).  All
three cases are examples of instrumentation for current or future
astroparticle experiments.  
\end{abstract}

\maketitle

\section{Motivation}

Recent generations of Field Programmable Gate Arrays (FPGAs) have
become incredibly powerful logic devices.  With companion
sensors and a data interface, they become valuable data collection
nodes for the physical sciences.  For certain applications, it has
become apparent that these logic devices themselves can serve as the
crucial processing elements of instrumentation electronics.
Specific applications where this resource can be tapped include, but
are not limited to:

\begin{itemize}
\item compact form factor (complete system with very small footprint)
\item ultra low-power compared with discrete systems of comparable functionality
\item input-output (I/O) limited systems
\item low Electromagnetic Interference (EMI)
\end{itemize}

Specific examples presented here use the Xilinx Spartan-3 FPGA, a
low-cost and flexible device generously made available through the
Xilinx University Program~\cite{XUP}.  High performance instrumentation
is increasingly a marriage between custom Application Specific
Integrated Circuits (ASICs) and high performance FPGAs, capable of
processing and transmitting the data collected.  In certain
applications, the external circuitry simplifies dramatically, and much
of the critical functionality can be provided by the FPGA itself.

Three examples are presented here as solutions to power, size and
noise constraints of current and planned neutrino detection
experiments.  While specific to these fields, the techniques are
generally applicable to measurement requirements in other disciplines.

\section{ANITA Discriminator}

\begin{figure}[htbp]
\epsfysize=3.5in
\epsfbox{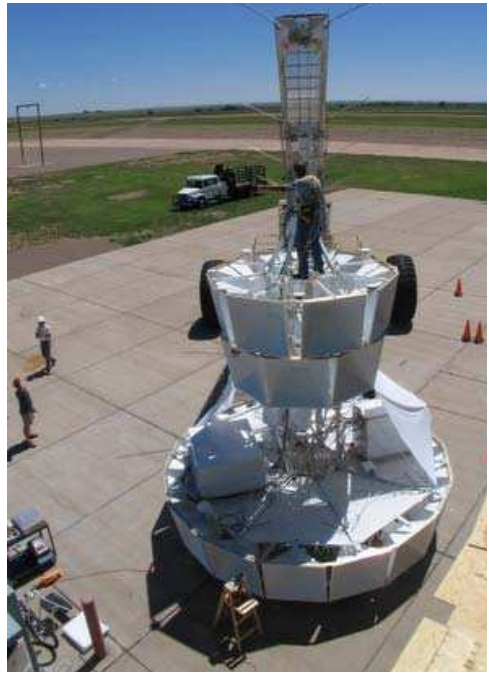}
\caption{\it Photograph of the ANITA engineering payload preparing to
launch.  Antarctic flight is scheduled for December 2006.
\label{EM} }
\end{figure}

The Antarctic Impulsive Transient Antenna (ANITA)~\cite{ANITA}\cite{SNIC}
experiment is a balloon-borne payload targeted to make the first
observation of the flux of ultra-high energy neutrinos, which must
exist due to the presence of the high-energy flux of cosmic ray
protons.  ANITA uses the generation of impulsive radio signals in the
Antarctic ice, produced by neutrino interactions at very high energy,
as the detection method.  A photograph of the ANITA payload preparing
for its engineering flight in New Mexico is seen in Fig.~\ref{EM}.

\begin{figure}[htbp]
\epsfxsize=3.2in
\epsfbox{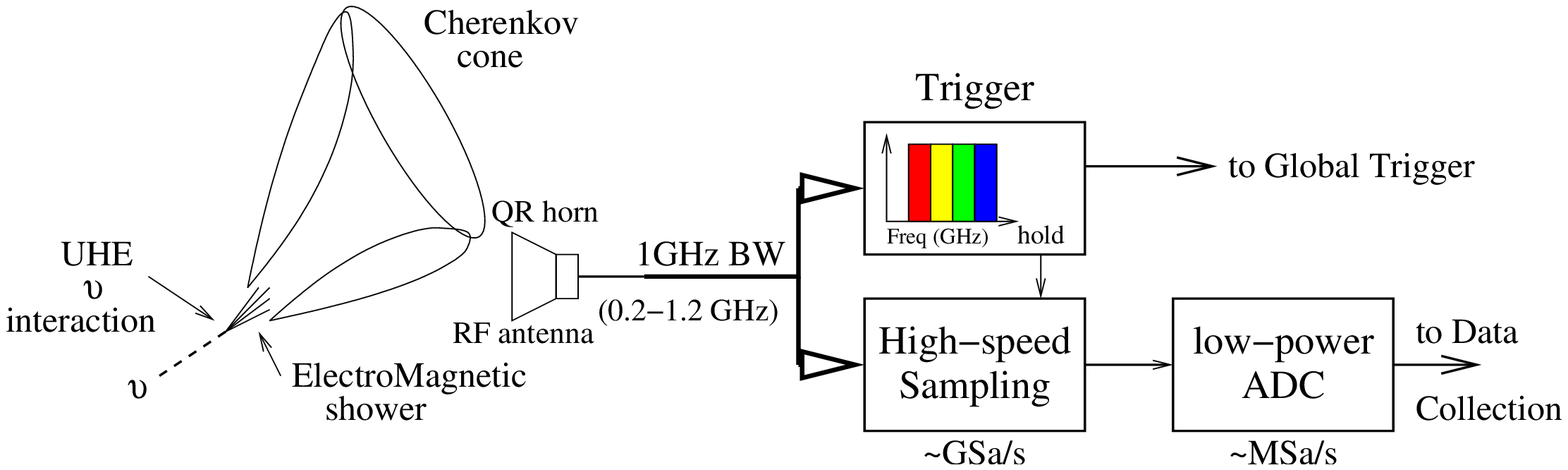}
\caption{\it In order to minimize the power required, signals from the
antennas are split into analog sampling and trigger paths.  To provide
trigger robustness, the full 1GHz bandwidth is split into 4 separate
frequency bands, which serve as separate trigger inputs.
\label{RF} }
\end{figure}

As a long-duration balloon-borne payload over the Antarctic continent,
power is a severe constraint on the instrumentation.  Particularly for
triggering and sampling many dozens of antennas over the full
bandwidth of 0.2-1.2 GHz, with greater than Nyquist sampling.  The
strategy adopted for ANITA is to split the signal as shown in
Fig.~\ref{RF} and use the trigger information to decide when the
analog samples~\cite{SPIE, LAB} should be held and digitized.  This scheme
can reduce the power required by two orders of magnitude or more.

\subsection{Triggering}

In order to distinguish background thermal noise from transient
radio-frequency signatures of neutrino interactions, the primary
trigger consists of 256 channels of high-speed discriminator.  A
discriminator is a combination of a fast comparator and a 1-shot circuit
to give a signal passing threshold a fixed width output for the
purposes of forming a well-defined signal coincidence.

In order to meet the physics requirements, thresholds for the trigger
must be pushed as far down into the thermal noise as can be sustained.
Studies indicate~\cite{SMEX} that this corresponds to single
filter-band trigger rates of 1-2MHz.  Operation at such a high rate near
threshold requires a comparator of exceedingly high gain-bandwidth
(GBW) product.  In addition, such comparators are likely to produce
significant amounts of EMI due to fast switching output signals.

\begin{figure}[htbp]
\epsfxsize=3.2in
\epsfbox{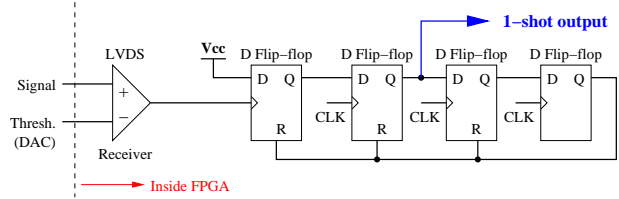}
\caption{\it Schematic of the basic discriminator circuit.
\label{input_ckt} }
\end{figure}

Within the FPGA considered, it turns out that with proper threshold
setting, the Low-Voltage Differential Signaling (LVDS)~\cite{LVDS}
comparator inputs provide exceptional GBW performance.  Internal logic
can be used to form an adjustable width output 1-shot circuit, immune
from ``runt'' (very narrow temporal pulse) generation, which is a
common problem in this type of discriminator logic.  This circuitry is
illustrated in Fig.~\ref{input_ckt}.  

Using a 250MHz {\tt CLK} signal, generated from the Digital Clock
Manager (DCM) as derived from a 33MHz (PCI clock) reference, the
discriminator output can be quantized in 4ns steps.  

Avoiding full signal swings external to the FPGA helps in two
significant ways: it reduces the power required by providing the input 
directly to the FPGA, and just as importantly, it reduces the EMI and
board-level noise induced by large, fast external switching signals.

\subsection{Implementation}

\begin{figure}[htbp]
\epsfxsize=3.2in
\epsfbox{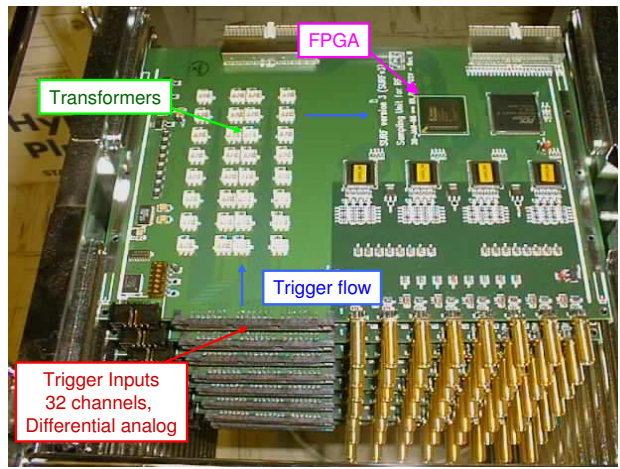}
\caption{\it Trigger signal flow across the Sampling Unit for Radio
Frequency board, with the 32 trigger signals terminating at the FPGA,
which serves as the discriminator.
\label{SURF} }
\end{figure}

The printed circuit board on which this discriminator is placed is
denoted the Sampling Unit for Radio Frequency (SURF) board and a photo
of the production board stack may be seen in Fig.~\ref{SURF}.  The
SURF form factor is 6U Eurocard and is designed to work in a
conduction-cooled compact-PCI crate.  

Conduction-cooling operation requires heat sinking (not shown) and
places emphasis on reducing component heat load.

Four filter-banded trigger signals for both polarizations of four
quad-ridge horn antennas are brought into the high density input
connector indicated.  These 32 differential analog signals are
converted to 50 $\Omega $ single-ended signals, where the other side
of the transformer is held at a reference potential which optimizes
the signal range to that of the LVDS receiver inputs.  Buried
microstrip lines are terminated in 50 $\Omega $ termination resistors
very close to the FPGA.  The trigger thresholds are set by applying a
voltage to the complementary input of the LVDS receiver, as seen in
Fig.~\ref{input_ckt}.  As this input is high-impedance, threshold setting
may be done with an array of compact, low-power, multi-channel output DACs.
Performance of this circuit is presented in detail in the next subsection.

An early prototype version of this board made use of eight 4-channel
high-speed comparators, which required additional space, power, and
produced significant noise.  With proper shielding, careful layout,
and working with small amplitude swings, the board level noise was
reduced drastically and appears largely dominated by ambient crate
noise.


\subsection{Performance}

\begin{figure}[htbp]
\epsfxsize=3.2in
\epsfbox{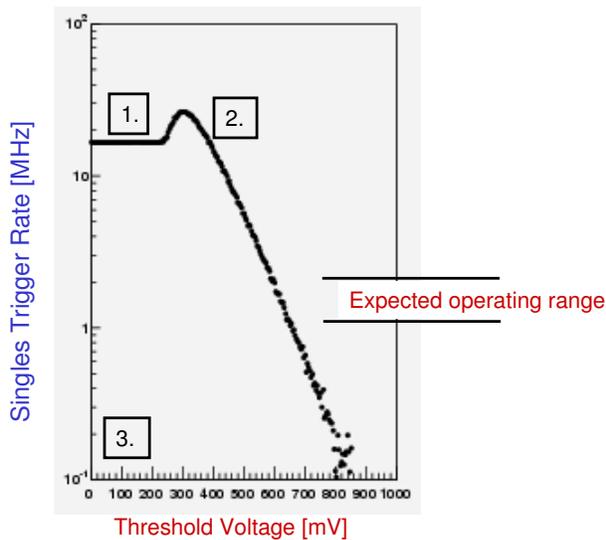}
\caption{\it Example of a threshold scan for a typical trigger
(amplified thermal noise) input signal.  Features of the curve shape,
as identified by the inset numbers, are described in the text.
\label{TC} }
\end{figure}

Performance of this discriminator was evaluated in a number of ways.
As a first step, a simple scan of count rate versus threshold for
amplified thermal noise, approximating the expected amplitude in
flight, was performed as shown in Fig.~\ref{TC}  

The curve follows expectations, considering three key features of the
12-ns output width 1-shot logic (adjustable in 4ns clock steps) and
its ``stuck on'' detect circuit.  These features are enumerated in the
figure and explained in detail below.

\begin{enumerate}
\item Saturation at 16MHz is the result of a ``stuck on'' detect
circuit, which is used to resolve the potential ambiguity in absolute
trigger rate.  That is, a measured rate could be due to the signal just
poking above threshold at a rate of 1MHz or only dropping below
threshold at a rate of 1MHz.  This circuit detects the latter
condition and enforces a minimum rate of 16MHz when the threshold is
too low.
\item The curve deviates from linear on this plot as the 12ns 1-shot
deadtime starts to become important.  At 8.3MHz (120ns average
period), this becomes a 10\% effect and eventually leads to the curve
rolling over, as can be seen.
\item Threshold voltage shown is on a relative scale with respect to a
voltage offset; only the relative voltage is important.
\end{enumerate}

\begin{figure}[htbp]
\epsfxsize=3.2in
\epsfbox{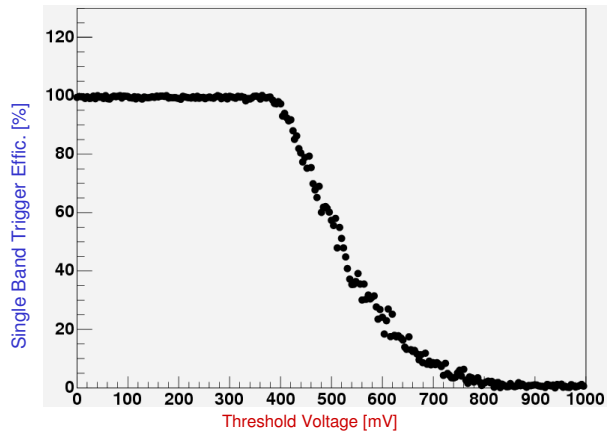}
\caption{\it Trigger efficiency scan as a function of threshold.
\label{ES} }
\end{figure}

Next a signal is embedded in the noise and the trigger efficiency is
studied as a function of threshold.  This scan is seen in
Fig.~\ref{ES} for a signal of about the minimum assumed in Monte Carlo
simulation of the ANITA sensitivity.  A trigger is determined if a
discriminator signal is coincident within 20ns of a reference strobe
from the pulse generator used.

\begin{figure}[htbp]
\epsfxsize=3.2in
\epsfbox{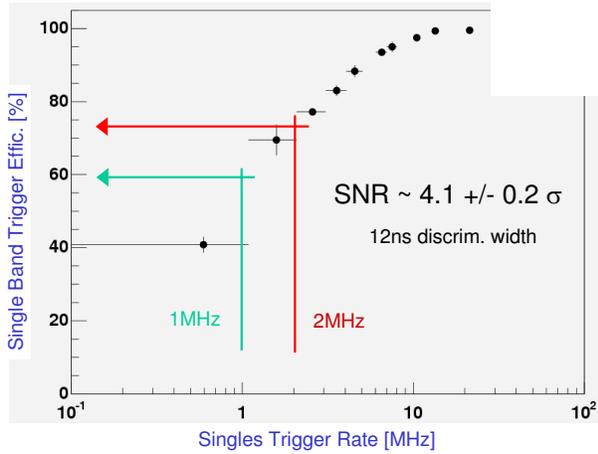}
\caption{\it Trigger efficiency versus singles rate.
\label{cmpfig} }
\end{figure}

Finally trigger efficiency is studied as a function of the threshold,
for the embedded signal with a Signal-to-Noise Ratio (SNR) of about
4.1$\sigma $ before filter banding.  As can be seen in
Fig.~\ref{cmpfig}, the efficiency is tuned as a function of the
tolerable singles rate in each channel.  

\begin{figure}[htbp]
\epsfxsize=3.2in
\epsfbox{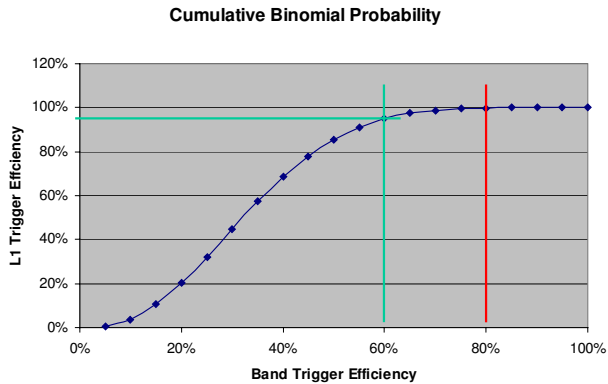}
\caption{\it Cumulative binomial probability of generating an overall
Antenna trigger versus the efficiency of individual band triggers. 
\label{3of8} }
\end{figure}

The limit on this rate is dictated by the overall accidental trigger
rate downstream, which we expect to be in the range indicated.  Since
the antenna trigger efficiency requires only a 3-of-8 coincidence, the
overall efficiency improves as can be seen in Fig.~\ref{3of8}.  

As seen in the figure, if an efficiency of about 80\% can be
maintained in each band, the overall Antenna-level trigger (``Level
1'' (L1)) efficiency is essentially unity.  Even a degradation to
60\% single band efficiency means an L1 efficiency of roughly 95\%.

\begin{figure}[htbp]
\epsfxsize=3.2in
\epsfbox{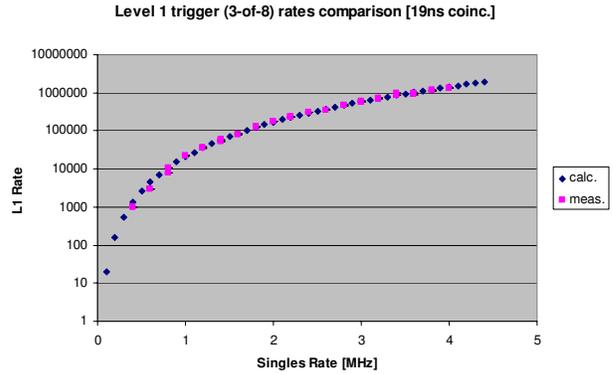}
\caption{\it Coincidence rate for accidental L1 triggers versus
singles rates with purely uncorrelated noise.  Overlaid are the
analytically calculable rates for completely random noise.  
\label{L1_rate} }
\end{figure}

In order to check the performance of the 1-shot circuit and
coincidence logic, a scan of L1 Rate was performed as a function of
single band rates for only thermal noise input (no signal).  The
resultant plot is seen in Fig.~\ref{L1_rate}.

Overlaid is the expected curve for the cumulative binomial probability
of purely random trigger band signals.  The two curves are in good
agreement for an effective coincidence window of 19ns, indicating a
5ns coincidence overlap requirement.

\section{HanoHano TDC and QDC}

\begin{figure}[htbp]
\epsfxsize=3.2in
\epsfbox{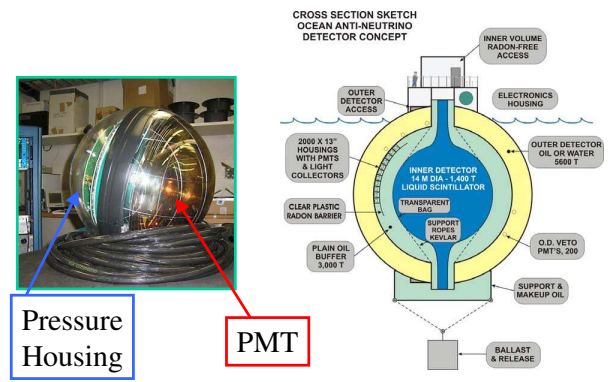}
\caption{\it Concept drawing of HanoHano (right) and a pixel of its
instrumentation: the Photo-Multiplier Tube (PMT) and its pressure housing
(left).
\label{HanoHano} }
\end{figure}

Detecting the much lower energy neutrinos due to nuclear fission or
fusion requires a detector of ultra-pure material.  The KamLAND
detector~\cite{KamLAND} has been very successful in measuring solar
neutrinos, as well as those from nearby power plants in Japan.
However, the naturally occurring radioactivity in the surrounding
mountains limit the sensitivity to possible neutrinos from decays deep
inside the earth.  To search for these and possible hidden
nuclear reactors, operation of such a detector deep in the ocean is
of interest.  A prototype of such a detector, designated
HanoHano~\cite{HanoHano}, may be seen in Fig.~\ref{HanoHano}.

The detector consists of a large volume of ultra-pure liquid
scintillator observed by a large array of Photo-Multiplier Tubes
(PMTs).  Deployment of such an extensive array far at sea is best done
without the requirements of a shore cable.  Therefore powering from a
battery system is very attractive if it can be made feasible.  Having a
single FPGA serve as Time-to-Digital Converter (TDC),
charge(Q)-to-Digital Converter (QDC) and data collection node, for a
large number of PMTs, represents a great power savings.

\begin{figure}[htbp]
\epsfxsize=3.2in
\epsfbox{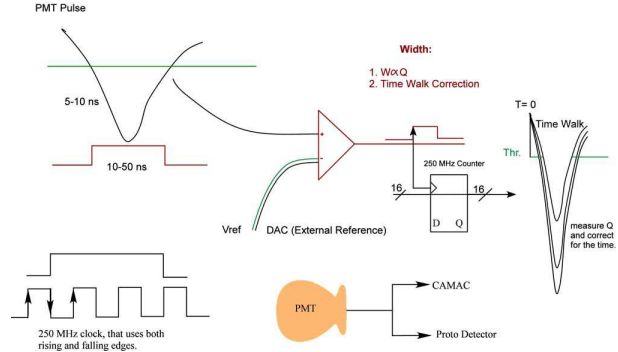}
\caption{\it Diagram of the low-power PMT readout processing.
\label{Hano_scheme} }
\end{figure}

Signal flow for the encoding of the PMT times in this readout can be
seen in Fig.~\ref{Hano_scheme}.  Again the LVDS receiver of the FPGA
is used as the comparator input.  The output of this comparator is
used to latch a time of the leading edge to measure time and the
trailing edge to measure the total charge Q, which is proportional to
the time-over-threshold.

Using both edges of a 250MHz clock, the expected ideal time resolution
should be

\begin{equation}
\sigma _{\rm T} = {T_{\rm clock} \over \sqrt{12}}
\end{equation}

where $T_{\rm clock}$ is the 2ns period when using both clock edges.
In this case, a resolution of about 0.6 ns is theoretically possible
and is explored below.

\subsection{Readout Implementation}

\begin{figure}[htbp]
\epsfxsize=3.2in
\epsfbox{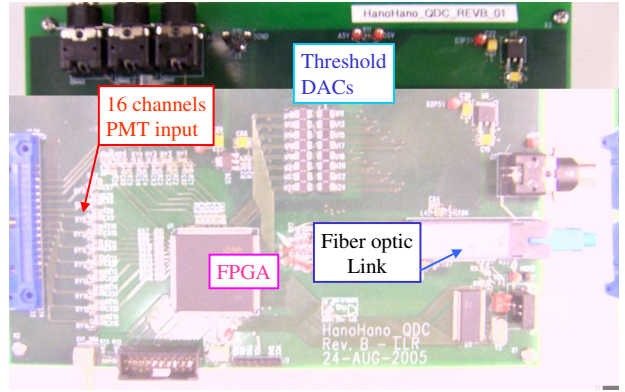}
\caption{\it Low-power HanoHano PMT Readout board.
\label{Hano_brd} }
\end{figure}

A prototype board for doing this readout can be seen in Fig.~\ref{Hano_brd}

A bundle of 16 PMT signals are brought onto the board via ribbon cable
at left and into the FPGA.  Continuous logging of the T and Q values
for each PMT hit is performed inside the FPGA and dumped into an
on-chip FIFO.  Data from this FIFO is collected over a fiber optic
link (at the right) to the central data acquisition computer.

\subsection{Test Results}

\begin{figure}[htbp]
\epsfxsize=3.2in
\epsfbox{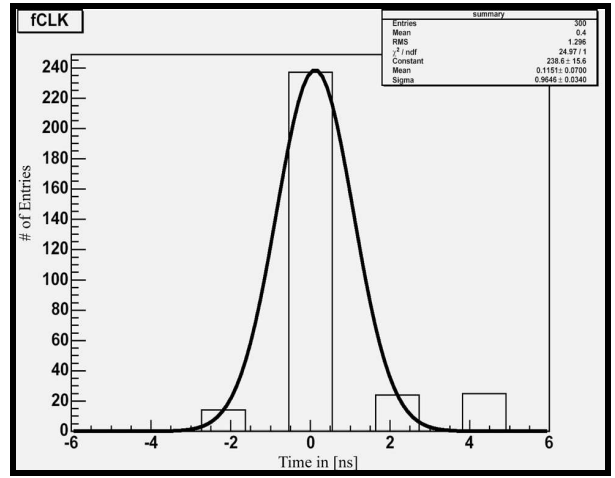}
\caption{\it FPGA TDC obtained timing resolution.
\label{Timing_resol} }
\end{figure}

The results from a simple Gray-code counter implementation of the hit
time recording may be seen in Fig.~\ref{Timing_resol}.  The sigma of
the distribution is about 1ns, but there are non-Gaussian tails.  In
this case the RMS resolution of about 1.3ns is more representative.  

While this is a factor of two worse than theoretical, it should be
pointed out that since the HanoHano PMTs are largely recording single
photons, and these large PMTs being employed have a Transit-Time-Spread
(TTS) of 3ns, the expected resolution will only degrade to 

\begin{equation}
\sqrt{(3ns)^2 + (1.3ns)^2} \simeq 3.3ns
\end{equation}

or by about 10\%.  Nevertheless, since high clock speeds and more
clever algorithms are possible, better timing will continue to be
pursed.

\section{SalSA ADC}

\begin{figure}[htbp]
\epsfxsize=3.2in
\epsfbox{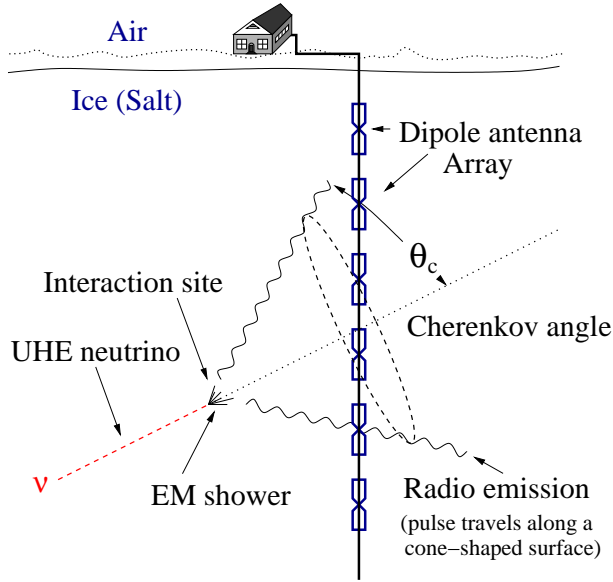}
\caption{\it Concept drawing for large volume radio neutrino telescope.
\label{SalSA} }
\end{figure}

Should the ANITA experiment confirm the expected flux of GZK
neutrinos, a next generation of terrestrial neutrino telescope will
exploit this source for both astronomy and particle physics.  An
example of such a detector is the proposed Salt-dome Shower Array
(SalSA)~\cite{SalSA}.  A schematic view of such a detector may be seen
in Fig.~\ref{SalSA}, where an array of drill holes into the salt (or
ice) are instrumented with strings of antennas.

As indicated in the figure, salt has the advantage of being 2.4 times
as dense as ice and of equally high radio transmission.  An earlier
study of low-cost readout was performed using short waveform sample
buffers~\cite{GEISER}.  Further studies have indicated that to optimize
neutrino flavor identification, improvements to the current generation of
low-power, compact Radio-Frequency sampling ASICs~\cite{SPIE,LAB} will
involve a readout capable of much longer waveform samples.  

\begin{figure}[htbp]
\epsfxsize=3.2in
\epsfbox{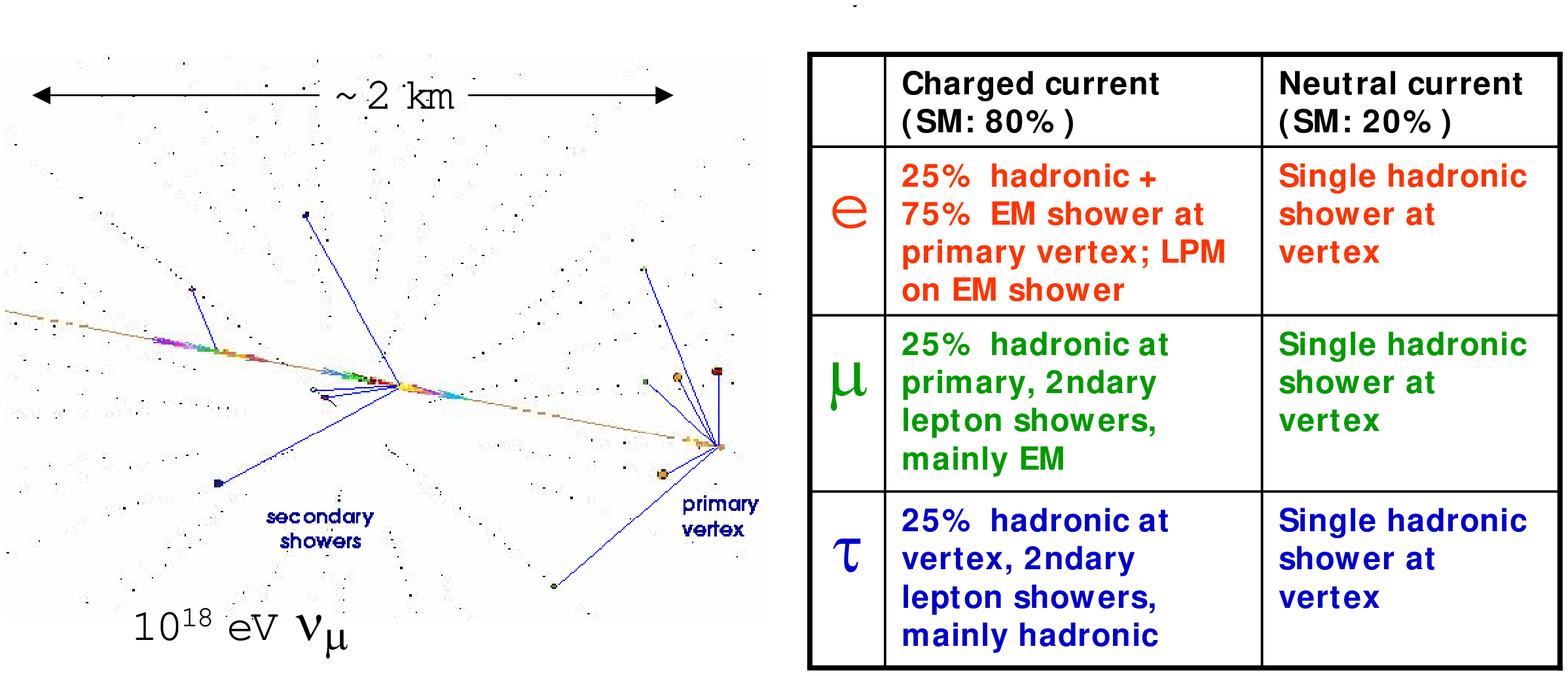}
\caption{\it Neutrino flavor identification through the observation of
the pattern of emission of secondary showers along the lepton path.
\label{nu_ID} }
\end{figure} 

This may be seen in Fig.~\ref{nu_ID}.  Particle identification is
possible by the flavor-specific pattern of subthreshold emissions
(secondary showers) along the lepton track.  However, to do so
requires the ability to store record lengths of order of the transit
time across the array in order to capture subthreshold signals.

\begin{figure}[htbp]
\epsfxsize=3.2in
\epsfbox{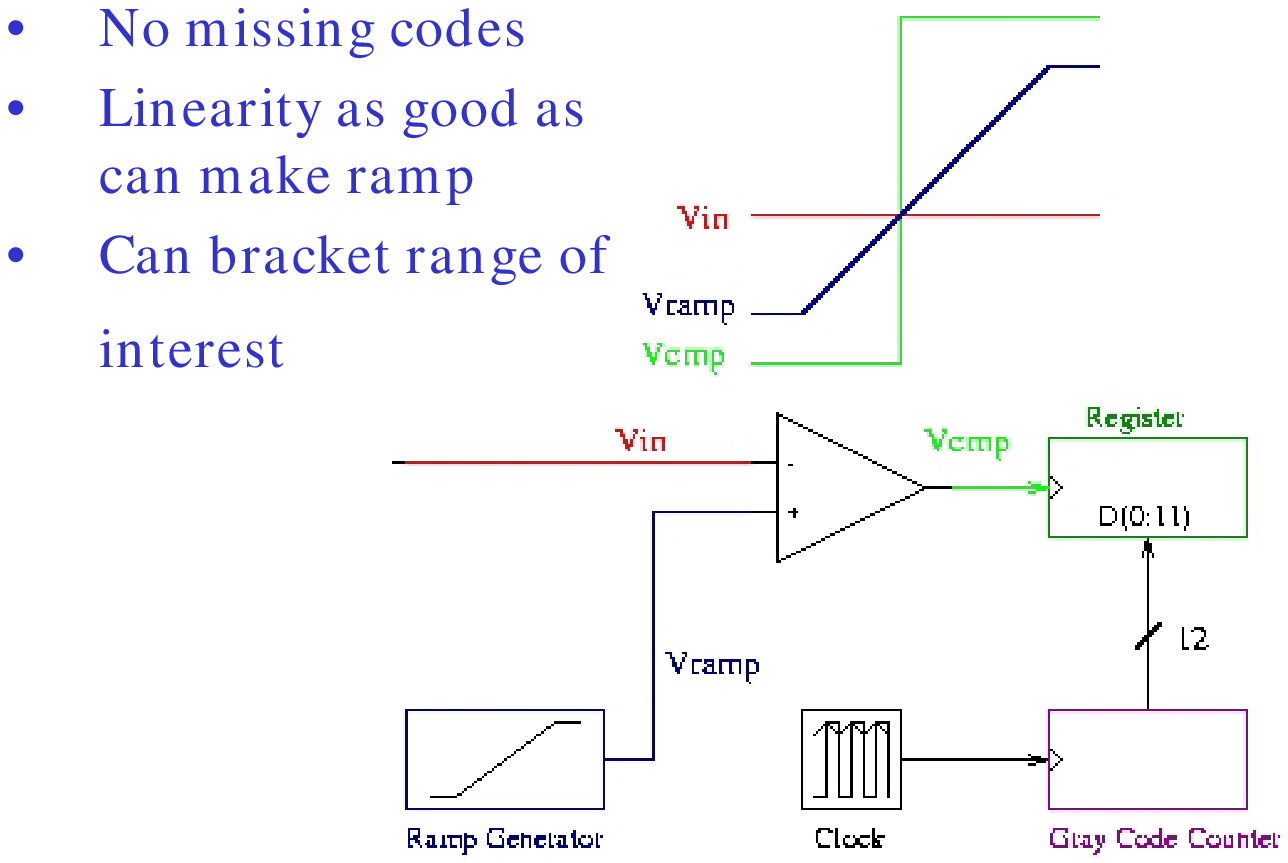}
\caption{\it Conversion of held samples using an external ramp and the FPGA.
\label{Wilk_ADC} }
\end{figure}

A proposal~\cite{BLAB1} was written to MOSIS~\cite{MOSIS} to fund the
fabrication of a test ASIC in a 0.25$\mu $m
CMOS process.  This device has 64k deep sampling, which
provides for up to 32$\mu $s of storage at 2GSa/s.  In order to
accommodate so many devices on a relatively small die, the conversion
of the storage samples using a Wilkinson-type ADC is performed by the
companion FPGA.  This scheme is illustrated in Fig.~\ref{Wilk_ADC}.

This scheme is quite general and with the use of an array of external
sample/hold circuits, or for rather slow speed signals, direct
conversion of analog signals with the FPGA may be performed.
Experience has shown that the performance of this type of encoding is
limited by the quality of the clock source and the linearity of the
ramp.  Fortunately the on-chip PLL provides for quite stable clock
sources and because the LVDS inputs are high impedance, a very simple
and yet high-quality ramp generator can be made with a current source,
a capacitor and a reset transistor.

\begin{figure}[htbp]
\epsfxsize=2.8in
\epsfbox{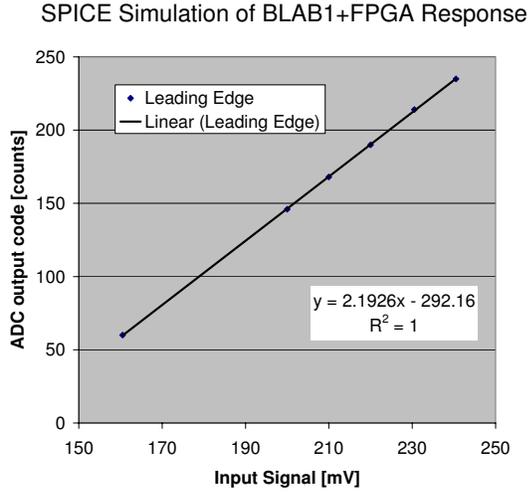}
\caption{\it Simulated performance of the ASIC (Buffered LABRADOR version 1
-- BLAB1) + FPGA analog recording.
\label{BLAB1} }
\end{figure}

Performance simulations have been done in SPICE and may be seen in
Fig.~\ref{BLAB1}.  The ASIC is currently in fabrication.

One advantage of performing the time encoding in programmable logic is
that the resolution can be tailored to the application.  A trade-off
can be made between speed and resolution in a highly flexible way.

\section{Future Prospects}

Future generations of FPGA will continue to improve in speed and
operate at reduced voltage levels.  For applications requiring good
timing, this will certainly be an improvement.  Current generations of
Xilinx already provide a stable 10GHz clock reference for the ``Rocket
I/O'' interface.  Having access to this clock would permit a
theoretical direct timing resolution of:

\begin{equation}
\sigma_{\rm timing} \simeq { {\rm T}_{\rm clock}\over \sqrt{12} } = 29ps
\end{equation}

However, at lower operating voltages, the limited dynamic range will
also prove a challenge.  Already these low rail voltages are an issue
in many mixed-signal applications.  

\section{Acknowledgements}

This work was enabled by the generous support of the ANITA
collaboration under the NASA Research Opportunities in Space Science
(ROSS) program, the HanoHano collaboration under the Center for
Excellence in Research in Ocean Science (CEROS) program, the MOSIS
Educational Program and the Xilinx University Program.


\setlength{\parindent}{0ex}


\begin{thebibliography}{99}


\bibitem{XUP} Xilinx University Program provides a donation program
(http://www.xilinx.com/univ) to higher education that has benefitted
all of the research projects described.

\bibitem{ANITA} P. Miocinovic {\it et al.} [ANITA Collaboration],
``Tuning into UHE neutrinos in Antarctica: The ANITA experiment'',
eConf {\bf C041213}, 2516 (2004); {\bf hep-ex/0503304}.

\bibitem{SNIC} G. Varner {\it et al.} [ANITA Collaboration],
``Detection of Ultra High Energy Neutrinos via Coherent Radio
Emission'', to appear in the Proceedings of the Stanford-Novosibirsk
Instrumentation Conference 2006, SLAC, April 2006.

\bibitem{SPIE} G. Varner {\it et al.}, ``Monolithic Multi-channel GSa/s
Transient Waveform Recorder for Measuring Radio Emissions from High
Energy Particle Cascades'', Proc. SPIE Int. Soc. Opt. Eng. 4858-31,
2003.

\bibitem{LAB} G. Varner, J. Cao, M. Wilcox and P. Gorham, {\it ``Large Analog
Bandwidth Recorder and Digitizer with Ordered Readout (LABRADOR)
ASIC.''}, manuscript in preparation for submission to
Nucl. Instr. Meth. {\bf A}, available online as {\bf physics/0509023}.

\bibitem{SMEX} ANITA Phase A proposal for the NASA
Small Explorer (SMEX) program, available online: 
www.phys.hawaii.edu/ \\
$\sim $anita/web/project/proposal/anitaprop.pdf

\bibitem{LVDS} Low-voltage differential signaling (LVDS) uses
high-speed analog circuit techniques to provide multi-gigabit data
transfers on copper interconnects.  It is defined under the Electronic
Industries Alliance (EIA)-644-1995 standard.

\bibitem{KamLAND} J.S. Ricol {\it et al.} (KamLAND Collaboration),
``KamLAND status and results'', EPS International Europhysics
Conference on High Energy Physics (HEP-EPS 2005), Lisbon, Portugal July
2005. PoS HEP2005:174, 2006.

\bibitem{HanoHano} S.~Dye {\it et al.} (HanoHano Collaboration),
``HanoHano: Detector Size and Sensitivity'', to appear in the proceedings of the
Neutrino Sciences 2005 - Geoneutrino Conference, Honolulu, Hawaii,
Dec. 2005.  Online link:
www.phys.hawaii.edu/$\sim $sdye/hnsc.html

\bibitem{SalSA} P.~Gorham {\it et al.}, ``Accelerator Measurements of
the Askaryan Effect in Rock Salt: A roadmap toward teraton underground
neutrino detectors.'', Phys. Rev. {\bf D 72} 023002 (2005).

\bibitem{GEISER} G. Varner {\it et al.}, ``A Giga-bit Ethernet
Instrument for SalSA Experiment Readout'', Nucl. Instr. Meth.  {\bf A
554} (2005) 437-443.

\bibitem{BLAB1} G. Varner, ``{\it Proposal to Fabricate:} First
Buffered LABRADOR (BLAB1) ASIC for High Speed Applications''.
Proposal awarded March, 2006.

\bibitem{MOSIS} The MOSIS multi-project wafer service is operated via the
University of Southern California, http://www.mosis.org





\end{thebibliography}
\end{document}